\RequirePackage{fix-cm}
\documentclass[twocolumn]{svjour3}          
\smartqed  
\usepackage{graphicx}
\usepackage{amsmath}
\usepackage{amssymb}
\usepackage{amsbsy}
\journalname{Journal of Computational Electronics}

\newcommand*{\affaddr}[1]{#1}
\newcommand*{\affmark}[1][*]{\textsuperscript{#1}}

\begin{document}

\title{Dips in high-order harmonics spectra from a subcycle-driven two-level system reflected in the negativity structure of the time-frequency Wigner function
}

\titlerunning{ }

\author{Seongjin Ahn\affmark[1] \and Andrey S. Moskalenko\affmark[1]}

\authorrunning{ }

\institute{Seongjin Ahn \at \email{seongjin.ahn@kaist.ac.kr}
\\
\\
Andrey S. Moskalenko \at \email{moskalenko@kaist.ac.kr}
\\
\\
\affaddr{\affmark[1]Department of Physics, KAIST, Daejeon 34141, Republic of Korea}
}

\date{Received: date / Accepted: date}

\maketitle

\begin{abstract}
We investigate high-order harmonics spectra radiated from a two-level model system driven by strong, ultrabroadband half- and single-cycle pulses, which are shorter than the inverse of the transition frequency. In this driving regime, the plateau in frequency spectra typical for radiation from strongly driven systems, has noticeable modulation in amplitude due to interference between waves of a same frequency and emitted at different time instants. Specifically, there is a characteristic `dips' structure at a set of frequencies in the radiation spectra, where the corresponding amplitudes are suppressed by several orders of magnitude. Understanding of this structure is required for applications such as generation of attosecond pulse, where number of composing modes and their relative phases are important. Therefore, we demonstrate a systematic way to find frequencies at which the dips are formed. To further illustrate the interference mechanism, we extract the phase information with the help of time-frequency distribution functions, namely the Husimi and Wigner functions. Especially, we found that the negativity structure of the Wigner function corresponds to each dip frequency and that the information regarding the type of interference is encoded in the pattern of the negative region of the Wigner function. Since such time-frequency Wigner function can actually be measured, we envisage utilizing its negativity structure to extract the phase information between radiation components emitted at time points within a subcycle time scale. This should provide an efficient tool for understanding and designing photonic applications, including short-wavelength coherent light sources.
\end{abstract}

\section{Introduction}\label{intro}

When a system is driven by light of central frequency $\omega$, not only the same mode is typically emitted but also modes of an integer multiple $N$ of $\omega$ may arise, called $N$-th order harmonics. When the driving field is strong enough, high orders of harmonics can be obtained, which have a series of applications such as the attosecond pulse generation and short-wavelength coherent light sources in a tabletop setup. Especially, among various candidates, solid-state-based systems driven by appropriate fields have been proposed as a promising platform for being compact, energy-efficient and capable of producing stable extreme ultraviolet light with robustness to fluctuations in the parameters of the driving pulse \cite{ghimire2011observation,ghimire2019high}.

In order to utilize the full potential of high harmonics from soild-state systems, several attempts have been made to model the underlying microscopic details of the mechanisms. In particular, a multi-level model has been shown to be capable of explaining unique features including multiple plateaus of solid-state high harmonics spectra while providing a concise picture of the radiation processes \cite{ndabashimiye2016solid,wu2016multilevel}. Remarkably, it was shown that although there are multiple energy levels interacting with each other, a set of two-level systems was a key building block to understand the harmonics radiation from solid-state systems \cite{wu2016multilevel}.

High-order harmonics radiation from two-level systems driven by multi-cycle pulses has been studied extensively and utilized to explain the key features of typical experimental high harmonics spectra, including peaks at integer multiples of the fundamental frequency followed by a plateau which extends up to a sharp cutoff \cite{sundaram1990high,faria2002high}. In these works, the driving field has a well-defined central frequency and the envelope duration of few oscillation cycles. Then, since the fundamental frequency is unambiguous, frequencies of the harmonics are also well-defined. Furthermore, although there are some fluctuations at the plateau region due to the interference between wave trains of a same central frequency emitted at each half cycle, if there are several cycles so that the number of emission time instants is not small the corresponding contributions tend to average out. This leads to the formation of a plateau structure in the spectra.

In this work, we go beyond the few-cycle case to deal with half-cycle and single-cycle driving pulses, being ultrabroadband in the frequency domain. We consider the subcycle driving regime with respect to the driven system, in the sense that the pulse duration is much shorter than the characteristic time scale of the system, e.g. $T_{0}\equiv 2\pi/\omega_{0}$, for the case of a two-level system with the energy difference $\hbar\omega_{0}$ between the two levels. Under the term `ultrabroadband', we specifically mean that the bandwidth of driving field is comparable to its central frequency, so that the fundamental frequency mode and its harmonics are not precisely defined. Here, we say `high-order harmonics' for indicating radiations of frequencies much higher than the central frequency of the driving field. In this situation, since the number of cycles is not high enough to ignore the interference between radiation bursts for a given frequency emitted at different time points, the plateau is no longer flat but has a set of missing frequencies due to the corresponding events of destructive interference. The information on the spectral distribution and phase in the high harmonics spectra is essential for applications such as attosecond pulse generation \cite{paul2001observation}. It is important to understand interference schemes and quantitatively predict those missing frequencies in the high harmonics spectra. In order to clearly reveal formation mechanisms of such spectral structures, we performed a time-frequency analysis using the Husimi and Wigner functions \cite{praxmeyer2007time}, the latter of which we found to be particularly for this goal.

We begin with a basic description of the driving scheme of the system in Section \ref{sec:description-of-system}, and evaluation of the resulting radiation spectra in Section \ref{sec:origin-of-radiation}. Then missing frequencies in the spectra are identified quantitatively with the help of the Husimi function in Section \ref{sec:radiation-spectra-and-dips}, followed by a discussion on how to utilize the Wigner function to extract the information on the interference involving more than two radiation time points. We describe a found set of correspondences between the patterns of negative regions in the Wigner function and the interference mechanisms, explained in Section \ref{sec:negativity-of-wigner}. We conclude this work with a summary on our findings and remarks on possible applications in Section \ref{sec:conclusion}.

\section{Theoretical model}
\label{sec:description-of-system}

\subsection{Two-level system driven by an ultrashort pulse in the subcycle regime}
\label{sec:model-two-level-system-driven-by-an-ultrashort-pulse}

We consider an electronic system interacting with an ultrashort pulse of light in the subcycle driving regime. The size of interaction region is assumed to be small enough so that the system perceives the electric field of light as nearly homogeneous throughout the region, i.e. we consider a case where dipole approximation is valid. We are interested in a case where just two levels of the system are relevant for the electron dynamics. Let us denote the two eigenvectors of the unperturbed Hamiltonian corresponding to the states of the lower and upper energy levels as $|1\rangle$ and $|2\rangle$, respectively. Then one can write the total Hamiltonian in terms of Pauli operators $\hat{\sigma}_{3} = |2\rangle\langle 2 | - |1\rangle\langle 1|$ and $\hat{\sigma}_{1} = | 2 \rangle \langle 1 | + | 1 \rangle \langle 2 |$, along with, for completeness, $\hat{\sigma}_{2} = -i | 2 \rangle \langle 1 | + i| 1 \rangle \langle 2 |$. Firstly, the unperturbed Hamiltonian can be written, up to a constant energy offset, as
\begin{equation}\label{eq:H0}
	\hat{H}_{0} = \frac{\hbar\omega_{0}}{2}\hat{\sigma}_{3}
\end{equation}
with $\hbar$ being the reduced Planck constant and $\omega_{0}$ being the transition frequency between the two states. Secondly, the electric-dipole coupling Hamiltonian is given by
\begin{equation}\label{eq:interaction-hamiltonian-V}
	\hat{V}(t) = -\hat{\mathbf{d}}\cdot\mathbf{E}(t),
\end{equation}
where $\hat{\mathbf{d}} = q\hat{\mathbf{x}}$ is the dipole moment operator with electronic charge $q = -e$. This operator can be expressed as $\hat{\mathbf{d}} = \mathbf{d}_{0}\hat{\sigma}_{1}$ under the two-level approximation and assuming that $|1\rangle$ and $|2\rangle$ having opposite parity. Here, the dipole moment matrix element $\mathbf{d}_{0}$ is defined as $\mathbf{d}_{0} = \langle 2 | \hat{\mathbf{d}} | 1 \rangle \in \mathbb{R}^{3}$, exploiting the invariance of quantum states under multiplication by an arbitrary complex unit number. Then the interaction Hamiltonian, Eq. (\ref{eq:interaction-hamiltonian-V}), becomes
\begin{equation*}
	\hat{V}(t) = -{\mathbf{d}_{0}}\cdot\mathbf{E}(t) \hat{\sigma}_{1},
\end{equation*}
which can be further simplified introducing an effective amplitude $E_{d}$ and a temporal shape function $f(t)$ of 
\begin{equation*}
	\mathbf{e}_{d} \cdot \mathbf{E}(t) \equiv E_{d}f(t),
\end{equation*}
where $\mathbf{e}_{d} \equiv \mathbf{d}_{0}/d_{0}$ is a unit vector pointing along the direction of $\mathbf{d}_{0}$. Further, one can express the interaction Hamiltonian as $\hat{V}(t) = f(t)\hat{V}_{0}$ with
\begin{equation}\label{eq:V0}
	\hat{V}_{0} = - \hbar\Omega \hat{\sigma}_{1}.
\end{equation}
Here $\hbar\Omega \equiv d_{0} E_{d}$, where $\Omega$ is the Rabi frequency characterizing the interaction strength. The total Hamiltonian becomes  
\begin{equation}\label{eq:total-hamil}
	\hat{H}(t) = \hat{H}_{0} + f(t)\hat{V}_{0}.
\end{equation}
In this work, we consider two types of the pulse shapes, $f_{h}(t)$ and $f_{s}(t)$, namely half-cycle and single-cycle Gaussian pulses:
\begin{subequations}
	\begin{align}
		f_{h}(t) & = \exp{ \left[ - t^2 / \tau_{d}^2 \right] }, \label{eq:pulse-half-cycle-gaussian}\\
		f_{s}(t) & = t\;f_{h}(t). \label{eq:pulse-single-cycle-gaussian}
	\end{align}
\end{subequations}
Here $\tau_{d}$ is a duration of the pulse.

\subsection{Time evolution in terms of rotation}

\begin{figure*}
\centering
\includegraphics[width=\textwidth]{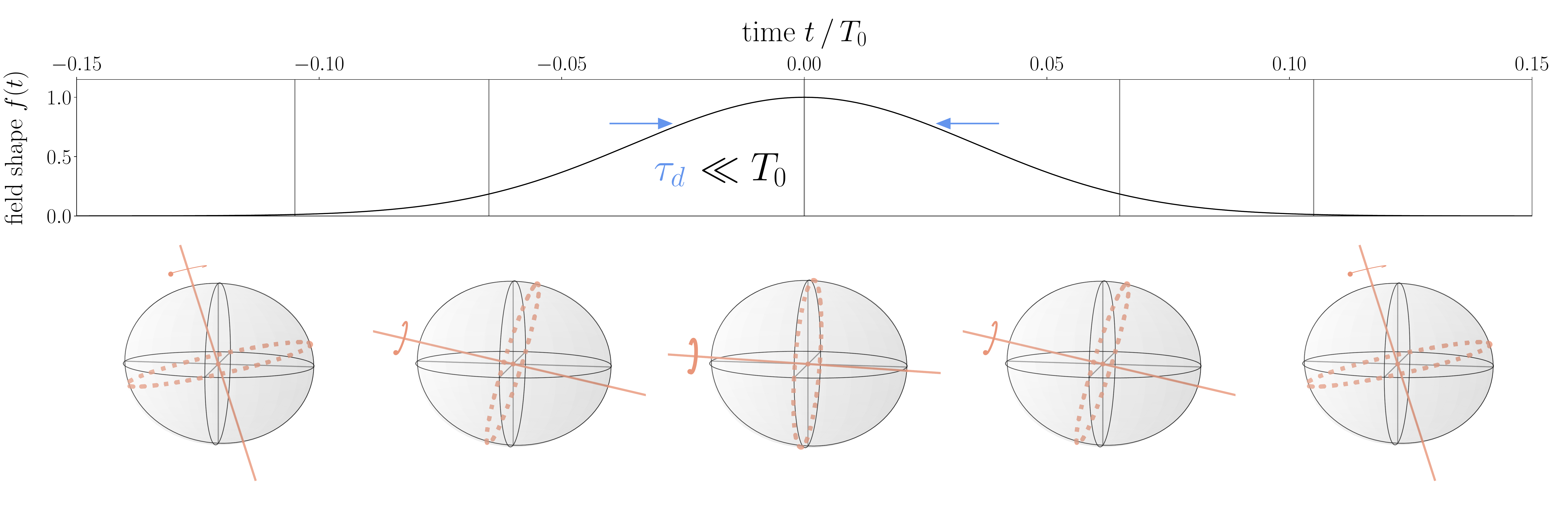}
\caption{Dynamics of the angular velocity vector of the Bloch vector. A half-cycle Gaussian pulse, Eq. (\ref{eq:pulse-half-cycle-gaussian}) is used for $f(t)$. Each sphere is represented for each time specified by the corresponding vertical line in the upper panel. Rotational axis $\mathbf{n}$ and angular speed $\Gamma$ in each sphere are determined by Eq. (\ref{eq:angular-velocity}), by the penetrating line and the thickness of the arrow around the line, respectively. Driving strength corresponds to the Rabi frequency $\Omega = 2\,\sqrt{\pi}/\tau_{d}$, pulse duration $\tau_{d}=T_{0}/20$.}
\label{fig:1-time-evolution-of-ang-freq-of-bloch-vector}
\end{figure*}

As a useful framework to understand the coherent dynamics of a pure quantum state, one can uniquely associate it with a point on a unit sphere, which is often called the Bloch sphere. This point corresponds to a 3-tuple $\pmb{\mathbf{\sigma}}$ comprising expectation values $\sigma_{j} := \langle \hat{\sigma}_{j} \rangle$, $j\in \{1,2,3\}$ of the Pauli operators, called Bloch vector. The time evolution of a Bloch vector $\pmb{\sigma}$ can be determined by evaluating the time derivative of each component following from the Schr\"{o}dinger equation $i\hbar|\dot{\psi}\rangle = \hat{H}|\psi\rangle$:
\begin{equation*}
	\dot{\sigma}_{j} = \langle [ \hat{\sigma}_{j}, \hat{H} ] \rangle / i\hbar.
\end{equation*}
Expressing the Hamiltonian as $\hat{H} = \hbar\Gamma_{k} \hat{\sigma}_{k}/2$ in the Einstein summation convention with $k\in{1,2,3}$, one arrives at the Bloch equation:
\begin{equation*}
	\dot{\pmb{\mathbf{\sigma}}} = \mathbf{\Gamma}\times\pmb{\mathbf{\sigma}}.
\end{equation*}
It prescribes, at a given time instant, an instantaneous rotation with an angular speed $\Gamma \equiv |\mathbf{\Gamma}|$ about the axis $\mathbf{n} \equiv \mathbf{\Gamma} / \Gamma$. For $\hat{H}$ given by Eq. (\ref{eq:total-hamil}), the instantaneous angular velocity vector reads
\begin{equation*}
	\mathbf{\Gamma} = \omega_{0} \mathbf{e}_{3} - 2\Omega f(t) \mathbf{e}_{1},
\end{equation*}
so that
\begin{subequations}\label{eq:angular-velocity}
	\begin{align}
		\Gamma(t) & = \sqrt{\omega_{0}^2 + \{2\Omega f(t)\}^2} \label{eq:angular-speed},\\
		\mathbf{n}(t) & = \cos{\theta} \mathbf{e}_{3} - \sin{\theta} \mathbf{e}_{1}
	\end{align}
\end{subequations}
where $\mathbf{e}_{j}$ for $j\in\{1,2,3\}$ are unit vectors in $\mathbb{R}^3$ corresponding to each direction. Also the inclination angle of the rotational axis $\theta \equiv \theta(t)$ is determined by 
\begin{equation}\label{eq:theta}
	\tan{\theta} = 2\Omega f(t) / \omega_{0}.
\end{equation}
In Fig. \ref{fig:1-time-evolution-of-ang-freq-of-bloch-vector}, we illustrate how the angular velocity $\mathbf{\Gamma}$ of the Bloch vector $\pmb{\mathbf{\sigma}}$ changes with time for a driving field given by Eq. (\ref{eq:pulse-half-cycle-gaussian}).

\subsection{Numerical computation of the time-evolving state}

To compute the time-dependent state vector and associated observables, we used the Crank-Nicolson method \cite{crank1947practical,bauer2017computational}. For components of the state vector $c_{n}(t) \equiv \langle n | \psi(t) \rangle$ in an ordered basis $\{ |2\rangle, |1\rangle \}$ formed by the eigenvectors of the unperturbed Hamiltonian $\hat{H}_{0}$ of Eq. (\ref{eq:H0}), the Schr\"{o}dinger equation reads
\begin{equation}\label{eq:tdse-matrix}
i\hbar\dot{\mathbf{c}} = \mathbf{H}\mathbf{c}
\end{equation}
with $\mathbf{c} \equiv [c_{2}, c_{1}]^{T}$ and $\mathbf{H}$ being a matrix representation of the total Hamiltonian $\hat{H}(t)$ in Eq. (\ref{eq:total-hamil}), given as
\begin{equation*}
\mathbf{H}(t)/\hbar = \
	\frac{1}{2}
	\begin{bmatrix}
		\omega_{0} & -2\Omega f(t)\\
		-2\Omega f(t) & -\omega_{0}
	\end{bmatrix}.
\end{equation*}

To solve Eq. (\ref{eq:tdse-matrix}), one may numerically evolve the initial state vector $\mathbf{c}(0)$ by a small time interval $\delta{t}$ repeatedly until reaching a desired final time. The matrix transformation corresponding to the time evolution, which should be unitary to preserve the norm of the state vector, can be constructed from a relation between the vector components $\mathbf{c}_{q} \equiv \mathbf{c}(t+q\delta{t})$ and the time evolution matrices $\mathbf{U}_{p,q} \equiv \mathbf{U}(t+p\delta{t},t+q\delta{t})$ for $p, q \in [0,1]$:
\begin{equation*}
	\mathbf{U}_{1/2,1}\mathbf{c}_{1} \
	= \mathbf{c}_{1/2} \
	= \mathbf{U}_{1/2,0} \mathbf{c}_{0}.
\end{equation*}
This basically says that $\mathbf{c}_{0}$ evolved by a half time step $\delta{t}/2$ forward is the same as $\mathbf{c}_{1}$ propagated by a half time step backward. Noting that the unitary matrices can be expanded for small enough $\delta{t}$, with $\mathbf{H}_{1/2} \equiv \mathbf{H}(t+\delta{t}/2)$,
\begin{equation*}
	\mathbf{U}_{1/2,{1/2}\mp{1/2}} \simeq \mathbf{1} \pm \frac{\mathbf{H}_{1/2}}{i\hbar} \frac{\delta{t}}{2} \equiv \mathbf{U}_{\pm}
\end{equation*}
one can finally derive the expression for the Crank-Nicolson evolution matrix,
\begin{equation*}
	\mathbf{U}^{CN} \equiv \mathbf{U}_{-}^{-1} \mathbf{U}_{+}.
\end{equation*}
Here applying the inverse matrix $\mathbf{U}_{-}^{-1}$ can be done either directly evaluating the matrix or by applying implicit algorithm such as lower-upper (LU) decomposition. It is known that the Crank-Nicolson method is stable in that the norm of the state vector remains constant being unconditional to the time step size ${\delta}t$ \cite{bauer2017computational}.

\section{Origin of classical radiation : dipole dynamics}\label{sec:origin-of-radiation}

Frequency spectra for the radiated classical light measured in far field can be evaluated by performing the Fourier transformation of a time-dependent dipole expectation value \cite{sundaram1990high}. In the considered case this expectation value is given by
\begin{equation*}
	\langle \hat{\mathbf{d}} \rangle = \mathbf{d}_{0} \sigma_{1} = d_{0}\sigma_{1}\mathbf{e}_{d},
\end{equation*}
where $\sigma_{1}(t) \equiv \langle \hat{\sigma}_{1} \rangle (t)$ is the dipole expectation value along the direction of dipole matrix element $d\equiv \mathbf{e}_{d} \cdot \langle \hat{\mathbf{d}} \rangle$ normalized by $d_{0}$,
\begin{equation*}
	d / d_{0} = \sigma_{1}.
\end{equation*}
the resulting frequency spectrum can be obtained in arbitrary units as:
\begin{equation}\label{eq:E-omega-abs-sq}
	\left| \tilde{E}(\omega) \right|^2 = \left| \ddot{d}_{\omega} \right|^{2},
\end{equation}
where $\ddot{d}_{\omega}$ is the Fourier transform of the second derivative of the dipole expectation value $\ddot{d}(t)$.

\begin{figure}
	\centering
	\includegraphics[width=0.45\textwidth]{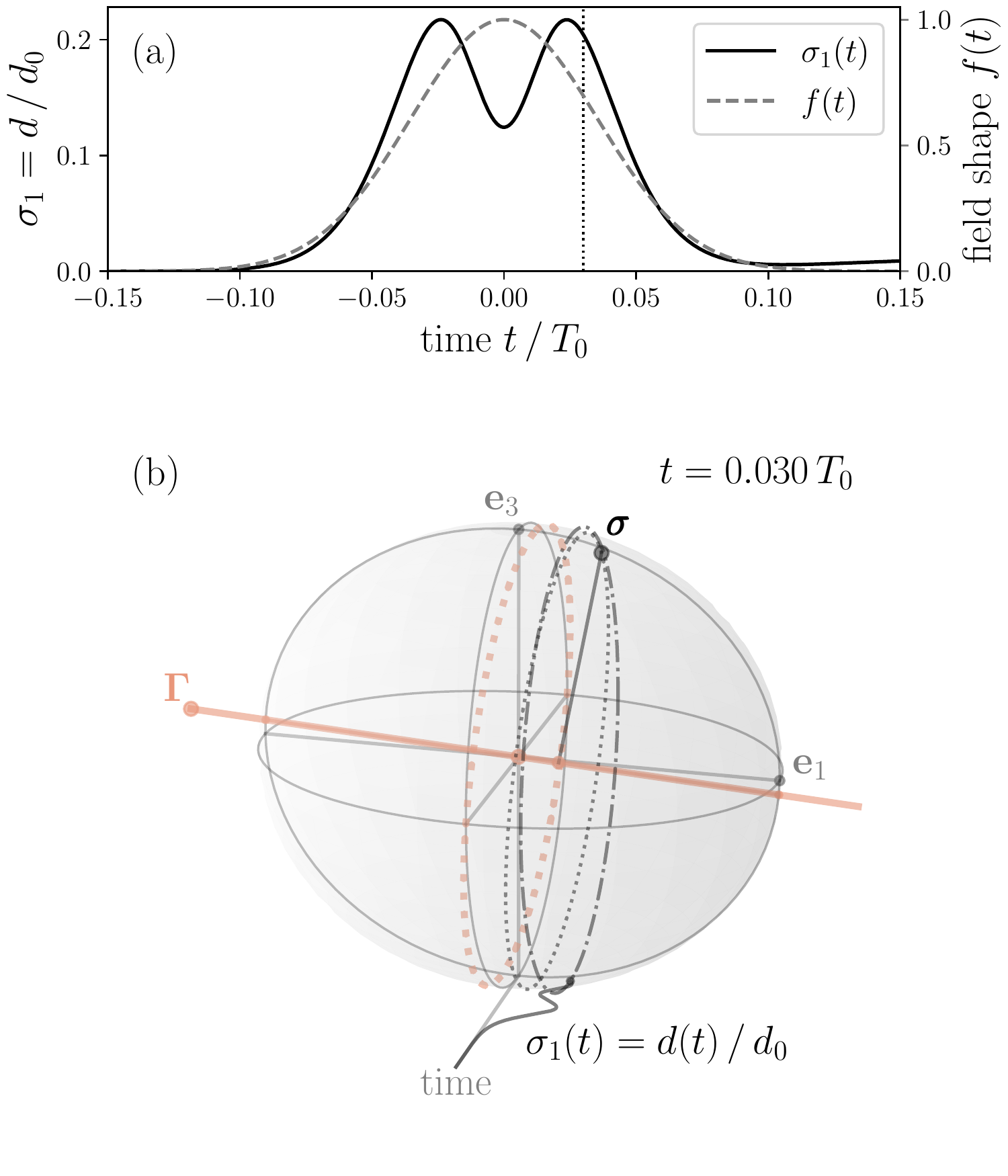}
	\caption{
		(a) Dipole expectation value $d(t)$ driven by a half-cycle Gaussian pulse of shape $f(t)$ given in Eq. (\ref{eq:pulse-half-cycle-gaussian}) with duration $\tau_{d} = T_{0}/20$ and interaction strength $\Omega = 2\,\sqrt{\pi}/\tau_{d}$. (b) A snapshot of Bloch sphere at an instant near the field maximum indicated by the dotted vertical line in (a). The points on the Bloch sphere represented by $\mathbf{e}_{1}$ and $\mathbf{e}_{3}$ correspond to eigenvectors of the Pauli operators $\hat{\sigma}_{1}$ and $\hat{\sigma}_{3}$, respectively. Projection of the Bloch vector $\pmb{\mathbf{\sigma}}$ onto the axis along $\mathbf{e}_{1}$ is plotted in the plane at the bottom of the sphere.
	}
	\label{fig:2}
\end{figure}

Fig. \ref{fig:2}(a) shows the dipole expectation value $d(t)$ in the case of subcycle driving by a half-cycle Gaussian pulse. It oscillates in the temporal vicinity of the field maximum at $t = 0$. This oscillation can be expected from the behavior of the Bloch vector depicted in Fig. \ref{fig:2}(b): oscillation in $\sigma_{1}$ comes from the rotation of the Bloch vector $\pmb{\mathbf{\sigma}}$ with $\mathbf{\Gamma}$ as its axis. Most of the oscillatory behavior occurs during the time interval when the angle $\theta$ of rotational axis $\mathbf{n}$ changes slowly compared to the angular speed $\Gamma \equiv |\mathbf{\Gamma}|$ near the field maximum. This time interval is of the order of the pulse duration $\tau_{d}$. Such oscillations can occur when the period of a full rotation near the field maximum $T_{\mathrm{peak}} = 2\pi / \Gamma_{\mathrm{peak}}$ fits into the pulse duration:
\begin{equation}\label{eq:T-peak-tau-d-condition}
	T_{\mathrm{peak}} \lesssim \tau_{d}.
\end{equation}
Since $\Gamma_{\mathrm{peak}} \sim 2\Omega$, as can be seen from Eq. (\ref{eq:angular-speed}) for $\Omega \gg \omega_{0}$, the condition in Eq. (\ref{eq:T-peak-tau-d-condition}) can be rewritten with respect to the interaction strength $\Omega$:
\begin{equation}\label{eq:regime-of-strong-Omega}
	\Omega \gtrsim \frac{1}{\tau_{d}}.
\end{equation}
In the regime specified by Eq. (\ref{eq:regime-of-strong-Omega}), frequencies higher than the frequency of the driving field are emitted effectively.

\section{Radiation spectra and `dip' structures}\label{sec:radiation-spectra-and-dips}

\subsection{Half-cycle driving pulse}

\begin{figure}
	\centering
	\includegraphics[width=0.45\textwidth]{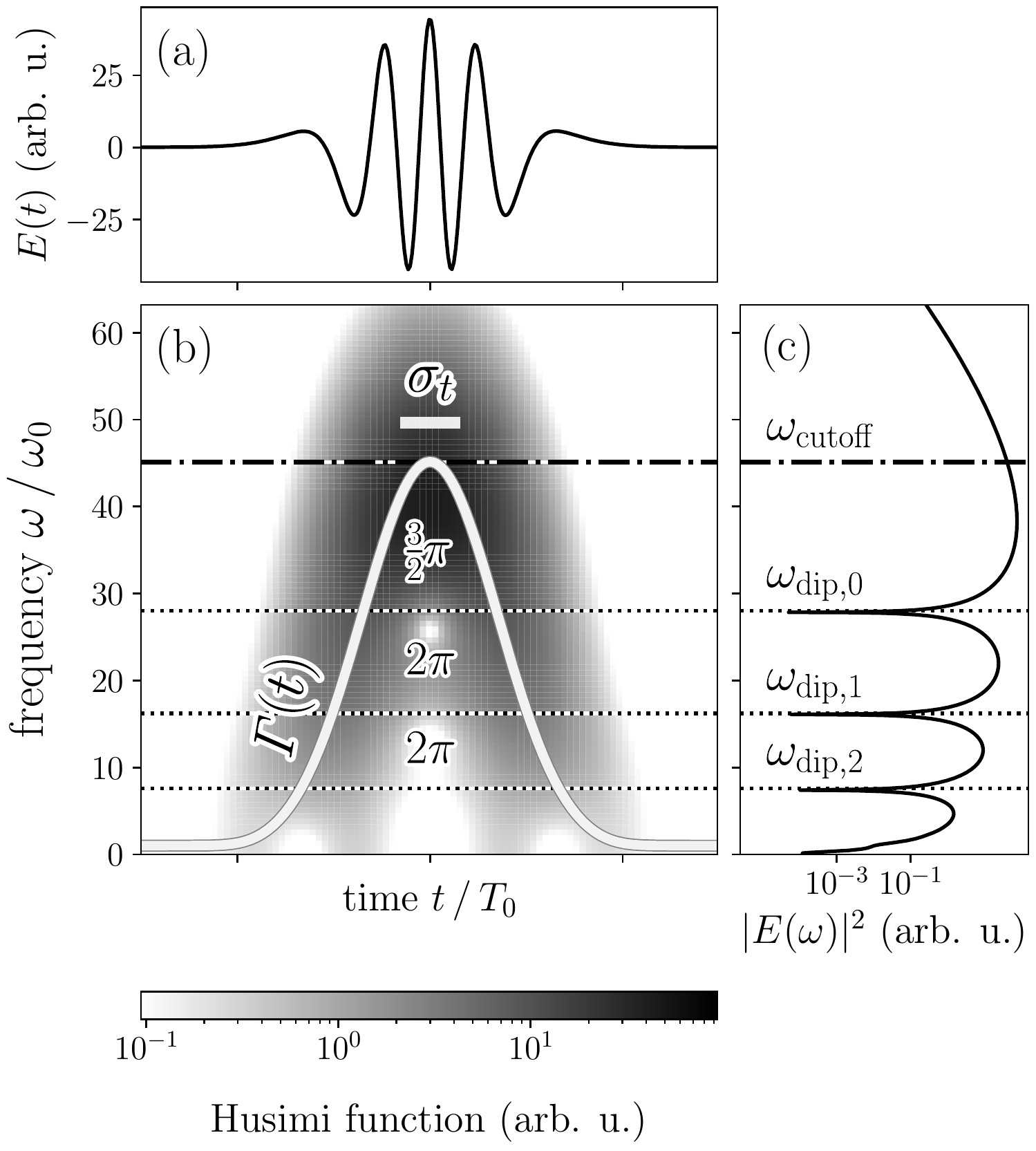}
	\caption{
		(a) Emitted electric field $E(t)$ from the system driven by a half-cycle Gaussian pulse with duration $\tau_{d} = \,T_{0}/20$ and interaction strength $\Omega = 4\,\sqrt{\pi}/\tau_{d}$. (b) Husimi function of the emitted electric field with time window of width $\sigma_{t} = \tau_{d}/(2\sqrt{2})$, indicated as a bar in the figure. Instantaneous angular speed $\Gamma(t)$ defined in Eq. (\ref{eq:angular-speed}) is overlaid on the Husimi function. Labels `$3\pi/2$', `$2\pi$', `$2\pi$' represent areas enclosed by $\Gamma(t)$ curve and respective dotted horizontal lines. These areas represent phase differences described in Eq. (\ref{eq:phase-dips}), which determine the positions of `dip' frequencies shown in (c). (c) Power spectrum for the emitted electric field shown in (a). Cutoff frequency $\omega_{\mathrm{cutoff}}$ beyond which the power spectrum monotonically declines coincides with the peak angular speed of the Bloch vector $\Gamma_{\mathrm{peak}} = \Gamma(0)$ shown in (b).
	}
	\label{fig:3}
\end{figure}

In Fig. \ref{fig:3}(a) we shoe the electric field $E(t)$ emitted from the system driven by a half-cycle Gaussian pulse with duration of $\tau_{d}=\,T_{0}/20$ and interaction strength of $\Omega = 4\,\sqrt{\pi}/\tau_{d}$, being in the regime of strong driving specified by Eq. (\ref{eq:regime-of-strong-Omega}). The emitted electric field is evaluated for the case of measurement in the far-field region, where it is proportional to the acceleration of dipole expectation value, namely $E(t) \propto \ddot{d}(t)$. Since the oscillation of $d(t)$ comes from the rotation of the Bloch vector, whose angular frequency $\Gamma(t)$ given by Eq. (\ref{eq:angular-speed}) is plotted in Fig. \ref{fig:3}(b), the emitted electric field in Fig. \ref{fig:3}(a) also changes its frequency with time. In other words, the emitted electric field is chirped. This can also be checked by evaluating time-dependent spectra \cite{eberly1977the} of the emitted electric field, constructred applying a convolution with a window function localized at a time instant of interest:
\begin{equation}\label{eq:husimi}
Q_{E}(t,\omega) = \left|\int_{\infty}^{\infty}{dt' E(t') G(t'-t;\sigma_{t}) e^{-i{\omega}t'}}\right|^2,
\end{equation}
where $G(t'-t;\sigma_{t})$ is centered at $t'=t$ and has width of $\sigma_{t}$. In the calculations of this work we used
\begin{equation}\label{eq:window-func}
	G(t;\sigma_{t}) = e^{-t^{2}/4\sigma_{t}^2} / \left(2\pi\sigma_{t}^2\right)^{1/4},
\end{equation}
when $Q_{E}$ is also called Husimi function \cite{grossman2008theoretical,praxmeyer2007time}. From the Husimi function in Fig. \ref{fig:3}(b), one can check that the dominant frequency radiated at each time instant basically follows the dynamics of the angular speed of the Bloch vector $\Gamma(t)$.

In Fig. \ref{fig:3}(c), total power spectra $|E(\omega)|^2$ is plotted according to Eq. (\ref{eq:E-omega-abs-sq}). As may be predicted from the connection of the dynamics of dipole expectation value and the rotation of the Bloch vector mentioned above, the highest possible frequency is given by the peak angular speed of the Bloch vector, as follows from Eq. (\ref{eq:angular-speed}),
\begin{equation}\label{eq:omega-cutoff}
	\omega_{\mathrm{cutoff}} = \Gamma_{\mathrm{peak}} \equiv \sqrt{\omega_{0}^{2} + (2\Omega f_{\mathrm{peak}})^2} \simeq 2\Omega f_{\mathrm{peak}}.
\end{equation}
Here $f_{\mathrm{peak}}$ is the peak value of the given pulse shape function $f(t)$. The last approximation comes from the condition of the subcycle driving regime $\tau_{d} \ll T_{0}$, implying $1/\tau_{d}\gg \omega_{0}$, and that of the strong driving regime satisfying Eq. (\ref{eq:regime-of-strong-Omega}).

Below the cutoff frequency there are frequencies at which the spectral density drops rapidly by a some orders of magnitude, forming `dip' structures. These features can be attributed to the interference between radiations of a same central frequency emitted at two different time instants. This fact can be anticipated analyzing Fig. \ref{fig:3}(b) and (c). For example, the highest dip frequency $\omega_{\mathrm{dip,0}}$ corresponds to two instants of time when the contributing wave trains are emitted. They are determined by intersections of the horizontal line at the level of $\omega{\mathrm{dip},0}$ with $\Gamma(t)$ in Fig. \ref{fig:3}(b). Although the existence of such interference effect has been known \cite{faria2002high}, here we deal with ultrabroadband pulses comprising just a half or a single optical cycle. Thus the modulation of the resulting spectra due to such interference effect is not merely smeared out to form a plateau typical for high-order harmonics spectra but leads to clear dip structures as demonstrated in Fig. \ref{fig:3}(c). Furthermore, we are able to report a quantitative identification of the dip frequencies for the case of half-cycle pulse, constituting a building block to understand the phenomena induced by few- and many-cycle pulses. To find these frequencies, one should look for a condition where radiation contributions with a given frequency below the cutoff are emitted at two different time points before and after the peak of the driving field and interfere destructively to yield a suppressed amplitude in the spectrum. If the angular speed of the rotation governing the emission frequency were constant $\Gamma(t)=\Gamma_{c}$ then the emitted light would basically be a monochromatic wave in a form of $\cos(\Gamma_{c}t)$, with no destructive interference but yielding a single peak at $\omega=\Gamma_{c}$ for the spectrum. In our case, $\Gamma(t)$ is not constant in time. For the two time points at which radiation wave trains with a given frequency of interest $\omega$ are emitted, denoted as $t_{\omega,-}$ and $t_{\omega,+}$, $\Gamma(t)$ satisfies $\Gamma(t_{\omega,\pm})=\omega$. $\Gamma(t) > \omega$ for the time interval between these time points $t \in (t_{\omega,-}, t_{\omega,+})$, as can be seen in Fig. \ref{fig:3}(b). Thus, during this time interval the emitted electric field shown in Fig. \ref{fig:3}(a) oscillates with instantaneous frequency $\Gamma(t)$ which is higher than $\omega$. The phase difference between the emitted electric field and that specified mode emitted at $t_{\omega,-}$ starts to accumulate as $\Gamma(t)dt - {\omega}dt$ during every passing time interval $dt$. Searching for a set of frequencies $\{ \omega_{\mathrm{dip},n} \}$ such that the accumulated phase difference makes the two contributions at each of these frequencies to interfere destructively and form a dip in the spectra, we obtain
\begin{equation}\label{eq:phase-dips}
	\phi_{\omega} \equiv \int_{t_{\omega,-}}^{t_{\omega,+}}{[\Gamma(t)-\omega]dt} = \phi_{0} + 2{\pi}n.
\end{equation}
Here $n$ is a non-negative integer and $\phi_{0} = 3\pi/2$. The dip frequencies $\omega_{\mathrm{dip},n}$ found by using this Eq. (\ref{eq:phase-dips}) well coincide with the actual locations of the dip structures in the spectrum shown in Fig. \ref{fig:3}(c).

\subsection{Single-cycle driving pulse}\label{sec:radiation:subsec:single-cycle-gaussian}

\begin{figure}
	\centering
	\includegraphics[width=0.45\textwidth]{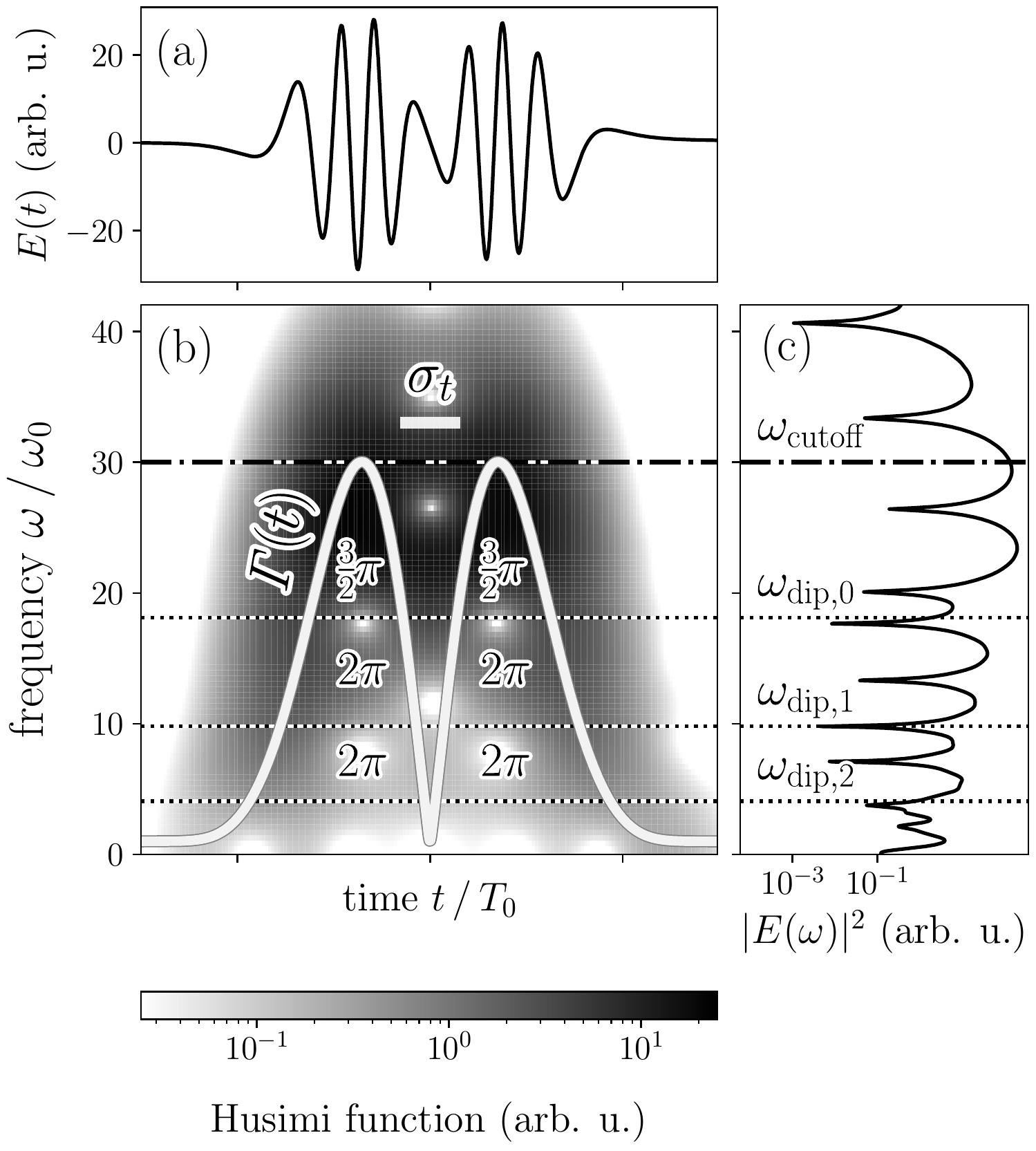}
	\caption{
		Same as Fig. \ref{fig:3} except that the pulse shape is single-cycle Gaussian, Eq. (\ref{eq:pulse-single-cycle-gaussian}) with duration $\tau_{d} =\,T_{0}/10$ and interaction strength $\Omega = 22/\tau_{d}$. A frequency $\omega$ between $\omega_{0}$ and $\omega_{\mathrm{cutoff}}$ has four time points at which contributions at that frequency $\omega$ are emitted as there are four intersections between the corresponding horizontal line at height of $\omega$ and the angular speed $\Gamma(t)$ of Bloch vector. The horizontal lines corresponding to dip frequencies $\omega_{\mathrm{dip},n}$ with $n\in\{0,1,2\}$ are drawn such that the areas enclosed by those horizontal lines and $\Gamma(t)$ are `$3\pi/2$', `$2\pi$', `$2\pi$' in the order from top to bottom as shown in (b). In addition to the indicated dip frequencies $\omega_{\mathrm{dip},n}$ given by Eq. ($\ref{eq:phase-dips}$) which was derived from an interference between two contributions of the same frequency both emitted within the same half-cycle, `intra-half-cycle' interference, there are additional dip structures which can be attributed to the phase differences between contributions originating from different half cycles, `inter-half-cycle' interference.
	}
	\label{fig:4:husimi-single-cycle}
\end{figure}

In Fig. \ref{fig:4:husimi-single-cycle} we show a time-dependent spectrum from the system driven by a single-cycle Gaussian pulse along with the emitted electric field and the total spectrum. The duration of the driving can be divided into two half cycles of the pulse, each of which has two time points where radiation of a given frequency below the cutoff is emitted, as shown in Fig. \ref{fig:4:husimi-single-cycle}(b). Since there are four contributions in total, more than one phase difference among them should be specified in order to characterize how they interfere to form a structure in the total spectrum.

In the total power spectrum plotted in Fig. \ref{fig:4:husimi-single-cycle}(c) there are a set of dip frequencies. We want to clarify the mechanism of their formation. Since we have identified dip frequencies for the case of half-cycle pulses as in Fig. \ref{fig:3} by using Eq. (\ref{eq:phase-dips}) and since a single-cycle pulse consists of half cycles, we can try to follow the same stratesy to identify at least some of the dip frequencies in the single-cycle driving case. Namely, we search for frequencies giving rise to contributions emitted twice within a half cycle with the phase difference acquiring the values given in Eq. (\ref{eq:phase-dips}). Thus these contributions interfere destructively. For the driving as in Fig. \ref{fig:4:husimi-single-cycle}, there are three such dip frequencies $\omega_{\mathrm{dip},n}$ with $n\in\{0,1,2\}$, leading to the phase differences of $3/2\pi + 2{\pi}n$ between the two respective time moments within each half-cycle. Note that the chosen single-cycle Gaussian pulses are symmetric with respect to $t=0$. Therefore the dip frequencies are the same for both half cycles, as evidenced by Fig. \ref{fig:4:husimi-single-cycle}(b). We observe that the frequencies found in this way agree well with the corresponding portions of the dips in the total spectrum Fig. \ref{fig:4:husimi-single-cycle}(c). Since the predicted dip frequencies, which comprise a part of the total set of dip locations, come from interferences between radiation contributions emitted \textit{within} a single half cycle, we may call them intra-half-cycle dip frequencies and denote as $\omega_{\mathrm{dip},n}^{\mathrm{intra}}$.


\section{Negativity structure of the Wigner function as an indicator for dip frequencies}\label{sec:negativity-of-wigner}

\subsection{Correspondence between negativity of the Wigner function and dip frequencies}

\begin{figure}
	\centering
	\includegraphics[width=0.49\textwidth]{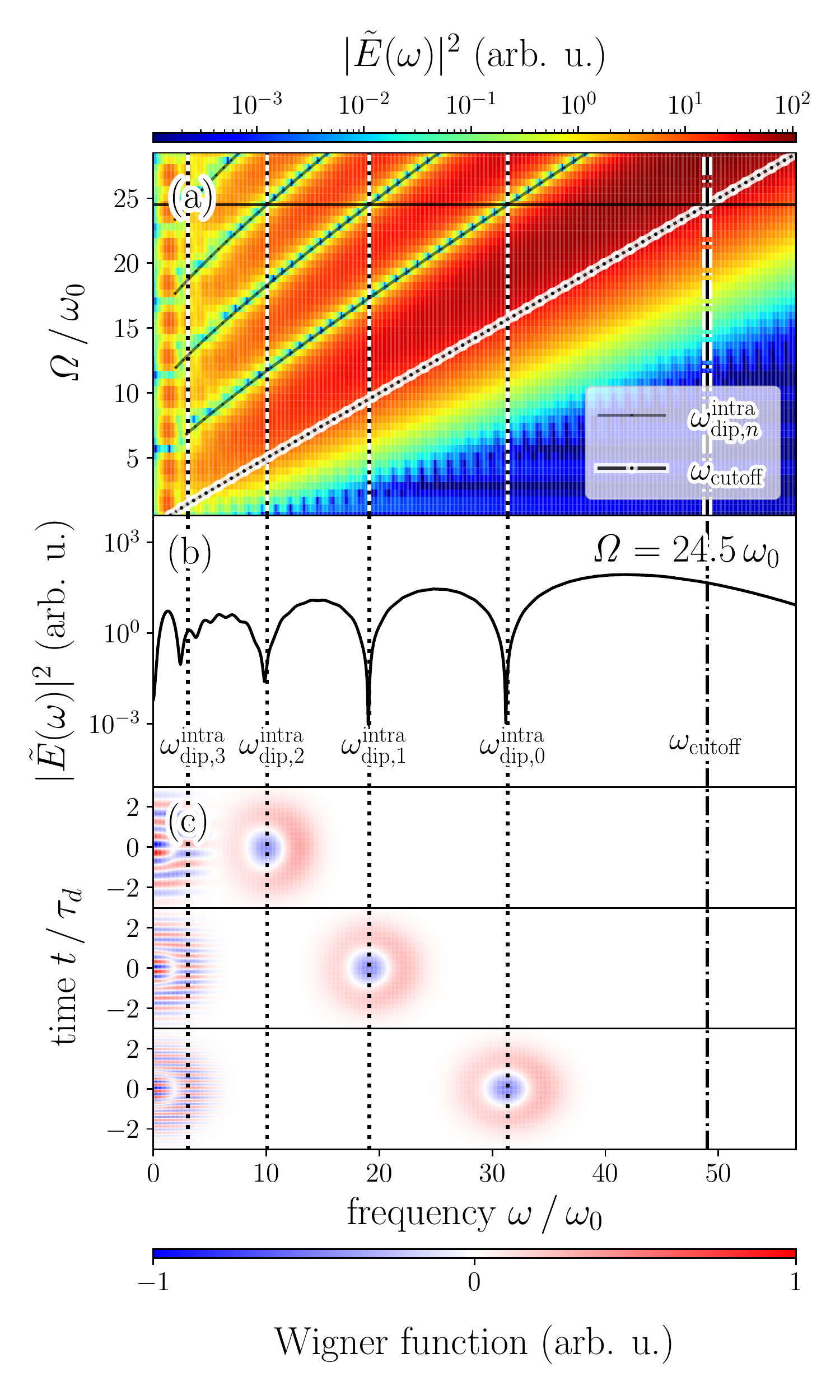}
	\caption{
		(a) Power spectra $|\tilde{E}(\omega)|^2$ of emitted electric field changing withe the driving strength $\Omega$. The the driving field shape is half-cycle Gaussian. Cutoff frequencies $\omega_{\mathrm{cutoff}}$ are proportional to the driving strength $\Omega$, as given by Eq. (\ref{eq:omega-cutoff}) (see diagonal lines). Frequencies at which dip structures appear are shown according to the prediction of Eq. (\ref{eq:phase-dips}). (b) Power spectrum at a selected driving strength $\Omega = 24.5\,\omega_{0}$ as indicated in (a) by a horizontal line. Dip frequencies $\omega_{\mathrm{dip},n}^{\mathrm{intra}}$ for $n\in\{0,1,2,3\}$ are shown. (c) Wigner functions for the emitted field with a window function of width $\sigma_{\omega} = 2\omega_{0}$ and centered in frequency domain at the dips frequencies of indices $n=2,1,0$, respectively.
	}
	\label{fig:5-spectra-wigner-funcs-half-cycle-gaussian}
\end{figure}

In Fig. \ref{fig:5-spectra-wigner-funcs-half-cycle-gaussian}(a), power spectra of the emitted electric field driven by a half-cycle Gaussian pulse are shown, in dependence on the driving strength $\Omega$. A particularly selected example of the spectra is depicted in Fig. \ref{fig:5-spectra-wigner-funcs-half-cycle-gaussian}(b) for the driving strength $\Omega = 24.5\,\omega_{0}$. The dip frequencies predicted by the intra-half-cycle interference picture as in Fig. \ref{fig:3} are indicated along with the cutoff frequencies following from Eq. (\ref{eq:omega-cutoff}). The spectra plotted in Fig. \ref{fig:5-spectra-wigner-funcs-half-cycle-gaussian}(a) and \ref{fig:5-spectra-wigner-funcs-half-cycle-gaussian}(b) are evaluated by the Fourier transformation, Eq. (\ref{eq:E-omega-abs-sq}), providing the amplitude at each frequency with the locations of dips. However, they do not provide, at least in a direct way, information on how such amplitudes are formed via the interference processes between radiations at the corresponding frequency. In order to obtain a signature which manifests the mechanism of such underlying interference processes, one needs to analyze the information related to the frequency of interest as well as the corresponding time-domain picture. The Husimi function which is an instance of the time-windowed Fourier transform, as in Eq. (\ref{eq:husimi}), has been used to analyze the emitted electric field, delivering time-dependent spectra which are local in time. Being useful to identify the time points at which the waves of the frequency of interest are emitted, however, the Husimi function may not be effective to combine signals from two or more distinct time points simultaneously, which is necessary to identify the interference mechanism. 

For this purpose, we used an inherently non-local transformation in time, provided by the Wigner function. Is is defined in the time-frequency domain as \cite{praxmeyer2007time}
\begin{equation*}
	W_{E}(t,\omega) = \int_{-\infty}^{\infty}{\frac{ds}{2\pi} E^{*}(t-s/2)e^{i{\omega}s} E(t+s/2)}.
\end{equation*}
By integrating the product of electric fields at different time points, extending the whole time domain, information from distinct time points is combined and encoded in the Wigner function. Then, to investigate the interference mechanism for each dip frequency, we multiply a window function to the emitted electric field in the frequency domain, centered at the frequency of interest with a certain width $\sigma_{\omega}$ as was done in e.g. \cite{kim2001wigner}. The frequency-windowed time-domain function is evaluated:
\begin{equation*}
	E_{\omega}(t) = \int_{-\infty}^{\infty}{\frac{d\omega'}{2\pi} \tilde{E}(\omega') G(\omega'-\omega;\sigma_\omega) e^{i{\omega'}t}},
\end{equation*}
where $\tilde{E}(\omega)$ is the Fourier transform of $E(t)$ and $G(\omega;\sigma_{\omega})$ takes the Gaussian form defined by Eq. (\ref{eq:window-func}).

The Wigner functions for the emitted electric field after applying the window transform centered at the dip frequencies are shown in Fig. \ref{fig:5-spectra-wigner-funcs-half-cycle-gaussian}(c) for the case of the half-cycle driving. We observe that at the time instant of the peak of the driving pulse, $t=0$, the Wigner functions show negative values centered close to each dip frequency. The negativity structures of the Wigner functions turn out to be indicators of dip frequencies originating in the underlying interference processes.

\begin{figure}
	\centering
	\includegraphics[width=0.49\textwidth]{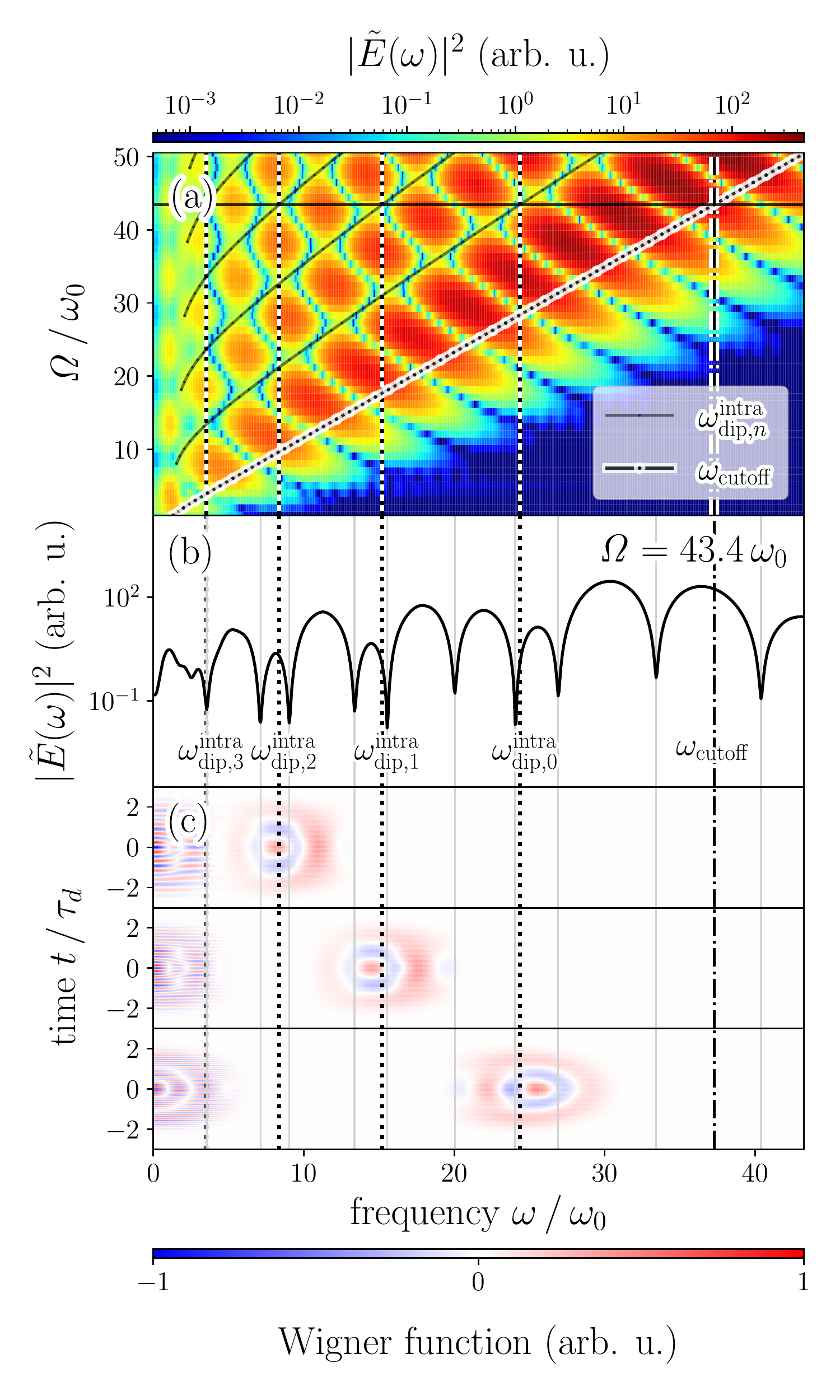}
	\caption{
		Same as Fig. \ref{fig:5-spectra-wigner-funcs-half-cycle-gaussian}, except that the driving pulse shape is single-cycle Gaussian. Also, among the dip frequencies, only those coming from the interference within each half-cycle, denoted as $\omega_{\mathrm{dip},n}^{\mathrm{intra}}$, are indicated following Eq. (\ref{eq:phase-dips}). The solid vertical lines in (b) and (c) indicate numerically obtained dip frequencies, by just detecting local minima of the spectra in (b).
	}
	\label{fig:6-spectra-wigner-funcs-single-cycle-gaussian}
\end{figure}

Such signatures in the Wigner function become more sophisticated for the case of the single-cycle driving. In Fig. \ref{fig:6-spectra-wigner-funcs-single-cycle-gaussian}(a), the corresponding power spectra are shown in dependence on driving strengths. Compared to the case of the half-cycle driving illustrated in Fig. \ref{fig:5-spectra-wigner-funcs-half-cycle-gaussian}(a), the structure of the power spectra induced by a single-cycle Gaussian pulse has a set of similarities including the cutoff and intra-half-cycle dip frequencies. One remarkable difference is that there are additional sets of dip frequencies that may not be attributed to intra-half-cycle interference, as pointed out in Section \ref{sec:radiation:subsec:single-cycle-gaussian}. The existence of such extra dip structures may be expected since, compared to the case of the half-cycle driving pulse there are now two half cycles to yield in total four emission instants for a given radiation frequency, not two. Thus, the outcome is determined by three phase differences between the corresponding four contributions. Since both half cycles have the same shape, phase difference between two the time points of emission within one half-cycle is fixed if that of the other side from $t=0$ is given. This leaves only two degrees of freedom, namely the intra- and inter-half-cycle phase differences, for controlling the interference scheme. As a result, after excluding the dip frequencies coming from the intra-half-cycle interference, $\{ \omega_{\mathrm{dip},n}^{\mathrm{intra}} \}$, we are left with the dip structures formed by the inter-half-cycle interference effect. For example, the power spectrum shown in Fig. \ref{fig:6-spectra-wigner-funcs-single-cycle-gaussian}(b) has a set of dip frequencies, four of which are due to the intra-half-cycle interference at $\omega_{\mathrm{dip},n}^{\mathrm{intra}}$ with $n\in\{0,1,2,3\}$, and the rest comes from the inter-half-cycle interference effects.

Let us consider the Wigner function shown in Fig. \ref{fig:6-spectra-wigner-funcs-single-cycle-gaussian}(c), with frequency windows centered at the three highest intra-half-cycle dip frequencies, in comparison with the corresponding total power spectra for a given driving strength $\Omega = 43.4\omega_{0}$ shown in Fig. \ref{fig:6-spectra-wigner-funcs-single-cycle-gaussian}(b). We can observe that the respective Wigner function has negative values at each dip frequency, whether it comes from the intra-half-cycle or inter-half-cycle interference. However, some parts of the negativity structure represent not just simple pits, but are connected to nearby frequencies, forming a donut-like pattern, which will be discussed in the following subsection.

\subsection{Correspondence between the negativity of the Wigner function and the avoided crossing of dip frequencies}

Investigating the picture of the power spectrum in dependence on driving strength for the case of the single-cycle driving, Fig. \ref{fig:6-spectra-wigner-funcs-single-cycle-gaussian}(a), we notice that there are several lines of intra-half-cycle dip frequencies with positive slopes in $\omega$-$\Omega$ space, along with another set of lines of dips with negative slopes following the inter-half-cycle dip frequencies. We would like to focus on the intersection points of those two sets of lines. An example of such an intersection point can be found in Fig. \ref{fig:6-spectra-wigner-funcs-single-cycle-gaussian}(a) at $\omega = \omega_{\mathrm{dip},2}^{\mathrm{inter}}$ and $\Omega = 43.4\,\omega_{0}$. However, a closer look at the vicinity of the alleged intersection point reveals that actually the dip lines do not cross. Instead, they repel each other, forming an `avoided crossing'. We observe in Fig. \ref{fig:6-spectra-wigner-funcs-single-cycle-gaussian}(b) that the precise position of the dip frequency near the avoided crossing at $(\omega,\Omega)=(\omega_{\mathrm{dip},2}^{\mathrm{intra}},\,43.4\,\omega_{0})$ in Fig. \ref{fig:6-spectra-wigner-funcs-single-cycle-gaussian}(a) is not the crossing point itself but is split into two nearby dips of which $\omega_{\mathrm{dip},2}^{\mathrm{intra}}$ is located around the middle.

Comparing the structure of the avoided crossings in Fig. \ref{fig:6-spectra-wigner-funcs-single-cycle-gaussian}(a) with the Wigner functions for frequency-windowed electric fields centered at the intra-half-cycle dip frequencies, we observe that each donut-shape pattern appears in the Wigner functions in Fig. \ref{fig:6-spectra-wigner-funcs-single-cycle-gaussian}(c) in the frequency range where an avoided-crossing structure is located in Fig. \ref{fig:6-spectra-wigner-funcs-single-cycle-gaussian}(a). Also, at the dip frequency relatively far from avoided crossings such as the one in between of $\omega_{\mathrm{dip},0}^{\mathrm{intra}}$ and $\omega_{\mathrm{dip},1}^{\mathrm{intra}}$ at $\Omega = 43.4\,\omega_{0}$, the Wigner function shows a simple negative spot, instead of a donut-shape pattern. We note that the connection between two different dip frequencies forming a donut shape found near an avoided crossing have not occurred for the case of the half-cycle driving even when the width of the window function in the frequency domain $\sigma_{\omega}$ has been extended to include more than one dip frequencies. This is consistent with the observation that there is no avoided crossing for the case of the half-cycle driving.

Based on the comparison between the power spectrum and Wigner functions, we are able to make the following correspondences: on the one hand, a simple pit of the negativity in the Wigner function represents a dip frequency relatively far from the region of avoided crossings, where only one of the two lines, corresponding to the intra- or inter-half-cycle dip frequencies, passes through. On the other hand, the negativity structure of the donut shape indicates an avoided crossing of the two lines. We point out that how a given dip structure is formed, i.e. whether via the intra-half-cycle interference or via the inter-half-cycle inteference, or contributed by both of them that occurs in the avoided crossing region, may not be possible to figure out by just looking at the total power spectrum as in Fig. \ref{fig:6-spectra-wigner-funcs-single-cycle-gaussian}(b). Rather, we can identify the interference scheme of each dip frequency by analyzing the spectra in the time-frequency domain with the help of the frequency-windowed Wigner function. Furthermore, since the Wigner function in the time-frequency domain can actually be measured experimentally \cite{praxmeyer2007time,praxmeyer2016direct}, the corresponding analysis should be feasible under realistic conditions.

\section{Conclusion}\label{sec:conclusion}

We investigated radiation spectra from a two-level system serving as a building block of the multi-level model, which in its turn can fully reproduce high-order harmonics radiation from solids. Especially in contrast to the conventional high harmonics generation, we considered driving fields that have only half- or single-cycle duration, so that the typical plateau in the radiation spectra starts to have clear dip structures. This implies that there is a set of missing frequencies with amplitudes suppressed by several orders of magnitude. The information on those missing modes is important for application of the high harmonics radiation in the context of the generation of attosecond pulses and design of short-wavelength coherent light sources.

We evaluated radiation spectra for half-/single-cycle pulses and observed a series of dips structures. Quantitative identification of dip frequencies at which such structures are formed was demonstrated. Three schemes of interference to form each dip structure were found, namely intra-half-cycle, inter-half-cycle, and their hybrid. Especially, we found that, in the driving strength dependent power spectra, there is a set of avoided crossings at the hybrid regions where the lines of dip frequencies formed by the intra- and inter-half-cycle interferences are supposed to cross each other but never do so. To further understand each interference scheme, we employed the time-frequency Wigner function which is an inherently non-local transformation, capable of collecting information on phase differences between radiation contributions emitted at distinct time moments during the pulse. From the analysis, we observed that the dip structures correspond to the negative-valued regions of the Wigner functions. Furthermore, we figured out that it can be identified from the negativity pattern of the Wigner function whether a given dip structure had originated from the interplay of both the intra- and inter-half-cycle interferences, or had been contributed by only one of them. Since such time-frequency Wigner function can actually be measured, we envisage utilizing the negativity structure of the Wigner function towards extracting the information on the relative phases of the radiation components emitted at the subcycle time scale during the pulse.


%

\section*{Declaration}
\subsection*{Funding}
This research was supported by the National Research Foundation of Korea (NRF) grant funded by the Korea government (MSIT) (2020R1A2C1008500).

\subsection*{Conflict of interest}
The authors declare that they have no conflict of interest.

\subsection*{Availability of data and material}
The datasets generated during and/or analysed during the current study are available from the corresponding author on reasonable request.



\end{document}